\documentclass[%
 reprint,
superscriptaddress,
showpacs,
 amsmath,amssymb,
prc,
]{revtex4-1}

\usepackage{gensymb}
\usepackage{graphicx}
\usepackage{dcolumn}
\usepackage{bm}
\usepackage{hyperref}
\usepackage[mathlines]{lineno}
\usepackage{subfigure}

\hypersetup{
    pdfnewwindow=true,      
    colorlinks=true,       
    linkcolor=blue,          
    citecolor=blue,        
    filecolor=blue,      
    urlcolor=blue           
}

\usepackage{tikz}
\usetikzlibrary{positioning}
\usetikzlibrary{calc}

\newcommand{\bigo}{\mathcal{O}}

\newcommand{\dalphat}{\delta\alpha_\textrm{T}}

\newcommand{\dphit}{\delta\phi_\textrm{T}}
\newcommand{\pt}{p_\textrm{T}}
\newcommand{\qt}{\vec{q}_\textrm{T}}
\newcommand{\qtm}{q_\textrm{T}}
\newcommand{\dpt}{\delta \pt}

\newcommand{\dptv}{\delta\vec{p}_\textrm{T}}
\newcommand{\Dptv}{\Delta\vec{p}_\textrm{T}}

\newcommand{\lepton}{\ell}

\newcommand{\nucleon}{\textrm{N}}

\newcommand{\pF}{p_\textrm{F}}

\newcommand{\proton}{\textrm{p}}

\newcommand{\ptl}{\vec{p}_\textrm{T}^{\,\lepton^\prime}}
\newcommand{\ptlm}{\pt^{\lepton^\prime}}
\newcommand{\ptni}{\vec{p}_\textrm{T}^{\,\nucleon}}
\newcommand{\ptnim}{p_\textrm{T}^{\,\nucleon}}
\newcommand{\ptn}{\vec{p}_\textrm{T}^{\,\nucleon^\prime}}
\newcommand{\pni}{\vec{p}_{\nucleon}}
\newcommand{\ptnm}{\pt^{\nucleon^\prime}}

\newcommand{\tfsi}{\tau_\textrm{f}}

\newcommand{\wboson}{W}
\newcommand{\maqe}{M^\textrm{QE}_\textrm{A}}

\newcommand{\ox}{\affiliation{Department of Physics, Oxford University, Oxford, Oxfordshire, United Kingdom}}
\newcommand{\ral}{\affiliation{STFC Rutherford Appleton Laboratory, Harwell Oxford, Oxfordshire, United Kingdom}}
\newcommand{\ic}{\affiliation{Imperial College London, Department of Physics, London, United Kingdom}}
\newcommand{\uc}{\affiliation{University of Colorado at Boulder, Department of Physics, Boulder, Colorado, USA}}

\begin{document}

\preprint{XXX}

\title{
Measurement of  nuclear effects in neutrino interactions with minimal dependence on neutrino energy
}

\uc\ic\ox\ral

\author{X.-G.~Lu}\email{Xianguo.Lu@physics.ox.ac.uk}\ox
\author{L.~Pickering}\ic
\author{S.~Dolan}\ox
\author{G.~Barr}\ox
\author{D.~Coplowe}\ox
\author{Y.~Uchida}\ic
\author{D.~Wark}\ox\ral
\author{M.~O.~Wascko}\ic
\author{A.~Weber}\ox\ral
\author{T.~Yuan}\uc

\date{\today}

\begin{abstract}
We present a phenomenological study of nuclear effects in neutrino charged-current interactions, using transverse kinematic imbalances in  exclusive measurements. Novel observables with minimal dependence on neutrino energy are proposed  to study   quasielastic scattering, and especially    resonance production. They should be able to provide direct constraints on nuclear effects in neutrino- and antineutrino-nucleus interactions. 

\end{abstract}

\maketitle


\section{Introduction}

The study of neutrino interactions presents  unique challenges. Unlike for its electroweak counterpart,  the charged lepton, the    energy of a neutrino is generally difficult to measure. Existing accelerator technologies are able to  provide a neutrino beam with a well defined  direction, and yet the beam energy spectrum, which is of critical importance in neutrino oscillation analyses~\cite{Adams:2013qkq, Abe:2015zbg}, is not well known. With a nuclear target, where neutrinos interact with bound nucleons, uncertainties from various nuclear effects arise. With  accelerator neutrinos in the GeV regime, individual bound nucleons are resolved. Their momenta, due to  Fermi motion ranging up to $\pF\sim200$~MeV/$c$, need to be considered; the energy required to release one of them  (binding energy) causes further fluctuations in the initial kinematics.
Moreover, multinucleon correlations have recently been conjectured to contribute significantly to the measured cross sections in this energy regime, evoking a reexamination of the long-used impulse approximation~\cite{Martini:2010ex, Nieves:2011yp}. Because the hadronic final states are produced inside a nuclear medium---the nucleus containing nucleons undergoing Fermi motion---they experience final-state interactions (FSIs) before exiting and therefore  both their kinematics and identity can be altered. In response to this, the nucleus can be excited or it can break up, emitting nucleons, pions and photons---collectively known as \emph{nuclear emission}. These effects further modify the final states from the \emph{basic} neutrino-nucleon interaction, which could have already been biased by multinucleon correlations, and introduce ambiguities, in both measurements and calculation, to efforts to identify the interaction channel. The highly convoluted nature of the problem is made manifest by the fact that these nuclear effects are all present in different channels for all nuclei except hydrogen~\cite{Lu:2015hea}.

Experimental efforts to understand charged-current (CC) neutrino-nucleus interactions, on which   spectrum measurements of accelerator neutrinos are based, have been focused on inclusive and semi-inclusive observables, such as total  cross sections and  lepton kinematics, which depend strongly on the neutrino energy and  cannot be directly compared across experiments because the neutrinos are produced in wideband  fluxes. To interpret the data, theories of  the basic  interactions and  nuclear effects have to be folded into approximated neutrino energy spectra. Therefore,   on one hand, accessing the nuclear effects precisely is  difficult, while on the other hand, understanding the nuclear effects helps determine the spectra. For example, Ref.~\cite{Lu:2015hea} shows that, with moderate FSIs, the calorimetric approach of neutrino energy reconstruction outperforms  kinematic methods; yet models of FSIs have not been extensively tested because of the lack of measurements of the final-state hadrons.  

In this work, we  propose to measure nuclear effects via the transverse kinematic imbalance between the charged lepton and the primary final-state hadron in an exclusive CC channel, such as   quasielastic scattering (QE) and  resonance production (RES). The measurement in RES is especially important for a better understanding of nuclear effects in both neutrino and antineutrino interactions.   Our discussion starts with the impulse approximation. We will first demonstrate the minimal dependence of hadronic variables on the neutrino energy, and describe the nuclear medium response to FSIs. Then we illustrate how these variables are affected by Fermi motion and FSIs. Subsequently the issue with respect to multinucleon correlations and an extension of the technique to electron-nucleus scattering will be  addressed briefly.

\section{Nuclear medium response} 

Consider a CC interaction on a nucleus. At the basic level  the neutrino $\nu$ 
interacts with a bound nucleon $\nucleon$ which then transits to another hadronic state $\nucleon^\prime$: 
\begin{align}
\nu+\nucleon\rightarrow\lepton^\prime+\nucleon^\prime,\label{eq:ccintele}
\end{align}
where $\lepton^\prime$ is the charged lepton. 
 In the rest frame of the nucleus, the bound nucleon is subject to Fermi motion with momentum $\vec{p}_\nucleon$,  and an energy-momentum $(\omega, \vec{q})$ carried by a virtual $\wboson$-boson ($\wboson^\ast$) is transferred to it as the neutrino scatters. In characterizing the interaction, the virtuality $Q^2\equiv q^2-\omega^2$ and the  invariant mass $W$ of $\nucleon^\prime$ are  used. Following energy-momentum conservation (the binding energy is neglected compared to the initial nucleon energy~\cite{binding}), the energy transfer reads
\begin{align}
\omega  &= \frac{Q^2+W^2-m_\nucleon^2+2\vec{q}\cdot\vec{p}_\nucleon}{2\sqrt{m^2_\nucleon+p^2_\nucleon}},\label{eq:polarization}\\
 &\sim \frac{Q^2+W^2-m_\nucleon^2}{2\sqrt{m^2_\nucleon+p^2_\nucleon}},\label{eq:etrans}
\end{align}
where $m_\nucleon$ is the mass  of $\nucleon$, and the last line follows from averaging out the direction of $\vec{p}_\nucleon$ in Eq.~\ref{eq:polarization}, which is a first order approximation because the \emph{polarization} term $\sim\vec{q}\cdot\vec{p_\nucleon}$ with opposite orientations of $\vec{p}_\nucleon$ for a give $\vec{q}$ does not exactly cancel as the $\wboson^\ast$-$\nucleon$ cross section is slightly different with the varying center-of-mass energy~\cite{polbias}. Below the deeply inelastic scattering (DIS) region---especially in QE and RES where $W$ equals the nucleon and dominantly the $\Delta(1232)$ resonance mass, respectively---the cross section   is suppressed when $Q$ is larger than the nucleon mass. The hadron momentum in these channels, as indicated by Eq.~\ref{eq:etrans},  ``saturates" if the  neutrino energy is above  the scale   $Q^2/2m_\nucleon\sim\bigo(0.5~\textrm{GeV})$ beyond which the charged lepton  retains most of the increase of the neutrino energy. 

\begin{figure}
\begin{center}
\includegraphics[width=\columnwidth]{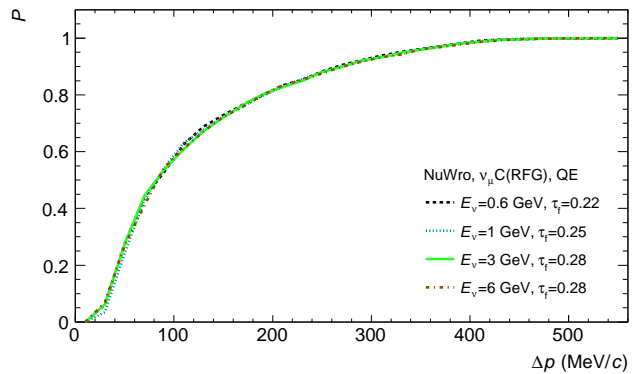}
\caption{Nuclear emission probability as a function of the in-medium momentum transfer, simulated by NuWro~\cite{Golan:2012wx} for $\nu_\mu$ CC QE on carbon---nuclear state modeled as  relativistic Fermi gas (RFG)~\cite{Moniz:1969sr}---at neutrino energy of 0.6, 1, 3 and 6~GeV.  Multinucleon correlations are ignored. The  in-medium interaction probability  $\tfsi$  (extracted from the simulation output throughout this work)  is shown in the legend. }\label{fig:pneqecmp}
\end{center}
\end{figure}

 Once the final state hadron $\nucleon^\prime$ is produced, it starts to propagate through the nuclear medium~\cite{noabs}. Under the assumption that the basic interaction (Eq.~\ref{eq:ccintele}) and the in-medium propagation are uncorrelated (i.e., are \emph{factorized}), the  momentum of $\nucleon^\prime$, which depends weakly on the neutrino energy,  completely determines the medium response, including the in-medium interaction probability $\tfsi$~\cite{tausf} and  the energy-momentum transfer $(\Delta E, \Delta\vec{p})$  to the medium (if $\nucleon^\prime$ decays inside the nucleus, the total effect of all decay products is considered). It is the latter that leads to nuclear excitation~\cite{deexcit} or break-up and consequently nuclear emission. The  nuclear emission probability, $P(\Delta E, \Delta \vec{p})$, correlates the medium response to the in-medium energy-momentum transfer~\cite{pnedef}. The factorization assumption suggests that $P(\Delta E, \Delta \vec{p})$ is independent of the neutrino energy $E_\nu$, which is consistent with the implementation in the NuWro~\cite{Golan:2012wx, gendep} simulation shown in Fig.~\ref{fig:pneqecmp}. In addition, as the neutrino energy increases, the predicted FSI strength  saturates, as is indicated by $\tfsi$ in the figure. 

\section{Single-transverse kinematic imbalance} 

\begin{figure}
\begin{center}
\includegraphics[width=0.6\columnwidth]{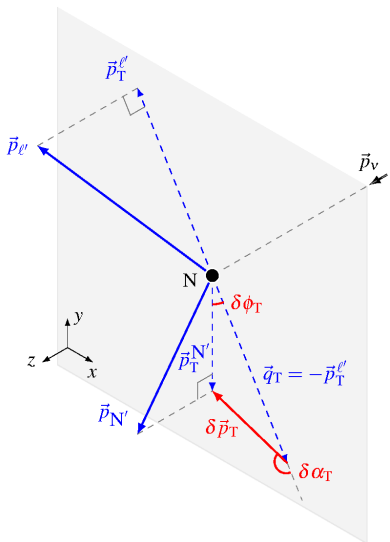}
\caption{Schematic illustration of the single-transverse kinematic imbalance---$\dphit$, $\dptv$ and $\dalphat$---defined in the plane transverse to the neutrino direction.}\label{fig:singleT}
\end{center}
\end{figure}

 To make a neutrino energy-independent measurement of nuclear effects, the in-medium energy-momentum transfer ($\Delta E$, $\Delta\vec{p}$) would be the ideal observable; this however is not experimentally accessible because of the unknown initial nucleon momentum  and the initially unknown neutrino energy. Instead, $\Delta\vec{p}$ can be directly inferred from the following single-transverse kinematic imbalance (Fig.~\ref{fig:singleT}):
 \begin{align}
\dptv&\equiv\ptl+\ptn,\label{eq:dpt}\\
\dalphat&\equiv\arccos\frac{-\ptl\cdot\dptv}{\ptlm\dpt},\label{eq:dalphat}
\end{align}
where $\ptl$ and $\ptn$ are the  projections of the extra-nucleus final-state  momenta transverse to the neutrino direction. In particular, $-\ptl=\qt$, the transverse component of $\vec{q}$.

 If the initial-state nucleon were static and free,     $\dpt$ would be  zero---a feature that is not possessed by other experimentally accessible  variables such as the final-state   momenta.
 If FSIs could be switched off,   $\dptv$ and  $\dalphat$ would be the transverse projection of $\vec{p}_\nucleon$ and of the angle between $\vec{p}_\nucleon$ and $\vec{q}$, respectively. Accordingly, to first approximation, the distribution of $\dptv$ would be  independent of the neutrino energy, and that of $\dalphat$ would be flat due to the isotropy of Fermi motion. 
   The FSI acceleration (deceleration) of the propagating $\nucleon^\prime$ adds in a smearing to $\dpt$ and  pushes $\dptv$ forward (backward) to $(-)\qt$, making $\dalphat\to0$ (180) degrees.

\begin{figure}
\begin{center}
\includegraphics[width=\columnwidth]{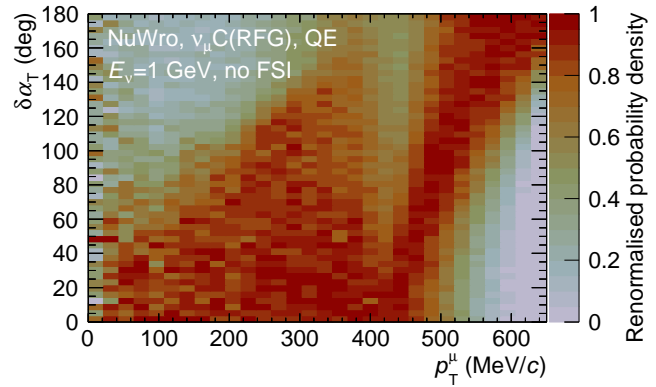}
\caption{Conditional probability density function of $\dalphat$ as a function of the muon $\pt$ without FSIs (each slice of $\pt^\mu$ is normalized in such a way that the maximum  is 1; the  renormalized density is shown on the $z$-axis), predicted by NuWro for $\nu_\mu$ CC QE on carbon (RFG) at neutrino energy of 1~GeV with FSIs switched off.  }\label{fig:2ddat}
\end{center}
\end{figure}

\begin{figure}
\begin{center}
\includegraphics[width=\columnwidth]{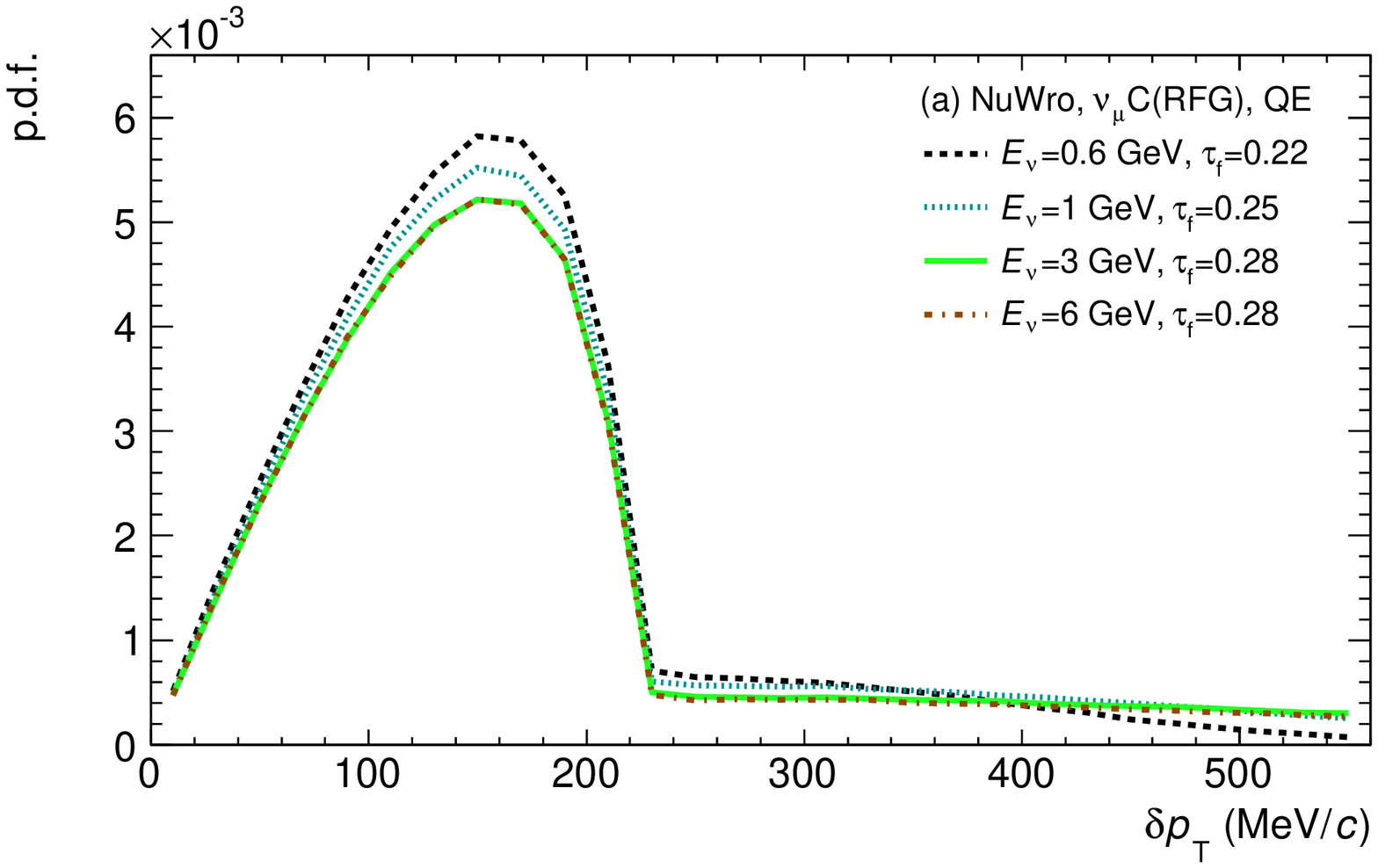}
\includegraphics[width=\columnwidth]{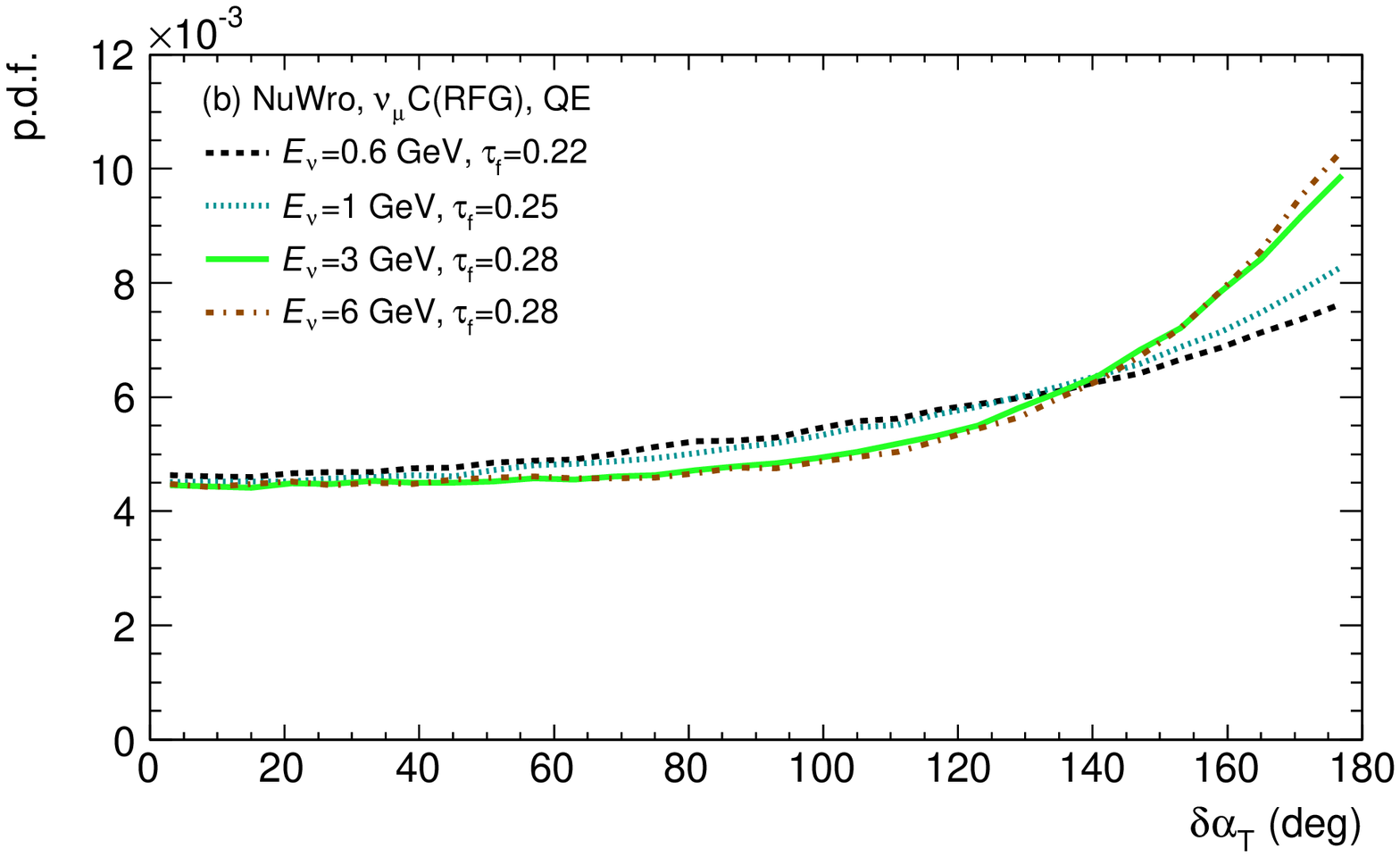}
\caption{Probability density function of $\dpt$ (upper) and $\dalphat$ (lower) predicted by NuWro for different neutrino energy.}\label{fig:enudptdalphat}
\end{center}
\end{figure}

 Second order effects that lead to the dependence on the neutrino energy include the previously discussed polarization (see text after Eq.~\ref{eq:polarization}), Pauli blocking, and the transverse projection. The combined effect determines the evolution of the $\dalphat$ distribution with $\ptlm$.  An example predicted by NuWro is shown in Fig.~\ref{fig:2ddat}.    At $\ptlm\lesssim\pF$, the cross section for  $\dalphat$ at 180 degrees is suppressed  in QE interactions due to Pauli blocking, which leads to a forward peak in the distribution of $\dalphat$ at small  $\ptlm$.  
As $\ptlm\to E_\nu$,  the cross section for $\dalphat$ at 0 degrees is suppressed by the conservation of the longitudinal momentum.   Even though the fractions of events in both  extremes of the $\ptlm$ spectrum change with the neutrino energy, they are insignificant for the few GeV neutrino interactions. As a result, the $\dpt$ and $\dalphat$ distributions   are largely independent of $E_\nu$, as is shown in Fig.~\ref{fig:enudptdalphat},  where the  evolution of the distributions with the neutrino energy  is dominated by variations in   the  strength of the FSIs.

The transverse momentum imbalance $\dpt$ has been used by the NOMAD experiment to enhance the purity of the selected QE~\cite{Lyubushkin:2008pe}, while the ``transverse boosting angle" $\dalphat$ is proposed here  for the first time.  
 Experimental data on $\dalphat$  will reveal the accelerating/decelerating nature of FSIs.  Its dependence on $\ptlm$, measured in a detector that has a low momentum threshold, will additionally provide    constraints on Pauli blocking.

\begin{figure}
\begin{center}
\includegraphics[width=\columnwidth]{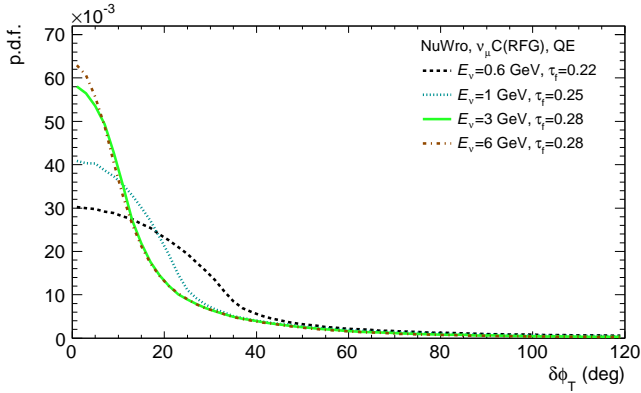}
\caption{Probability density function of $\dphit$.}\label{fig:enudphit}
\end{center}
\end{figure}

Besides the transverse momentum imbalance and boosting angle, another single-transverse variable can be defined (Fig.~\ref{fig:singleT}):
\begin{align}
\dphit&\equiv\arccos\frac{-\ptl\cdot\ptn}{\ptlm\ptnm}, \label{eq:dphit}
\end{align}
 which measures the  deflection of $\nucleon^\prime$ with respect to  $\vec{q}$  in the transverse plane.  If the initial-state nucleon were static and free,   $\dphit$  would be  zero; with nuclear effects, the deflection caused by $\Delta\vec{p}$ adds in a smearing to the initial distribution of $\dphit$ that is determined by $\vec{p}_\nucleon$. Experiments have measured the $\dphit$ distribution in QE-like events~\cite{Walton:2014esl} and used it to enhance the QE purity ~\cite{Lyubushkin:2008pe, Abe:2015oar}.  However,  the trigonometric relation illustrated by Fig.~\ref{fig:singleT} shows that $\dphit$ scales with $\dpt/\ptlm$ and therefore depends on the lepton kinematics which are  sensitive to the neutrino energy. The energy dependence of $\ptlm$ counteracts the FSI deflection and the uncertainties from the nuclear effects and   neutrino flux become  convolved. The distribution of $\dphit$ by NuWro is shown in Fig.~\ref{fig:enudphit} for different neutrino energies. In contrast to the expected evolution with the FSI strength, the distribution becomes narrower at higher energy because of the increase of $\ptlm$. This serves as an example of how the neutrino energy dependence can bias a measurement of nuclear effects.  Because of the    $\ptlm$ dependence, the single-transverse variables all suffer to some extent from a dependence on  the neutrino energy  even after  kinematic saturation is reached. Nevertheless, the study of nuclear effects can be performed  by restricting $\ptlm$. 

\section{Model predictions}

In the previous discussion, an equivalence is established between the nuclear effects in neutrino-nucleus interactions and the transverse kinematic imbalance. Initial and final-state effects can be directly observed via $\dptv$, as can be seen by rewriting Eq.~\ref{eq:dpt} into 
\begin{align}
\dptv&=\ptni-\Dptv, \label{eq:dptdef}
\end{align}
where $\pni$ is the momentum of the initial nucleon. In this section we present the latest predictions of the single-transverse variables. Interactions of neutrinos from the NuMI (on-axis) beam line~\cite{Anderson:1998zza} on a carbon target are simulated by  NuWro (version 11q)~\cite{Golan:2012wx} and GENIE (version 2.10.0) with the  hA FSI model~\cite{Andreopoulos:2009rq}. Since the neutrino energy is well above the saturation scale $\bigo(0.5~\textrm{GeV})$, the minimal energy dependence of the transverse kinematic imbalance applies. Interesting features of the implemented nuclear effects in the models are therefore maximally preserved and readily identified as shown below.

\begin{figure}
\begin{center}
\includegraphics[width=\columnwidth]{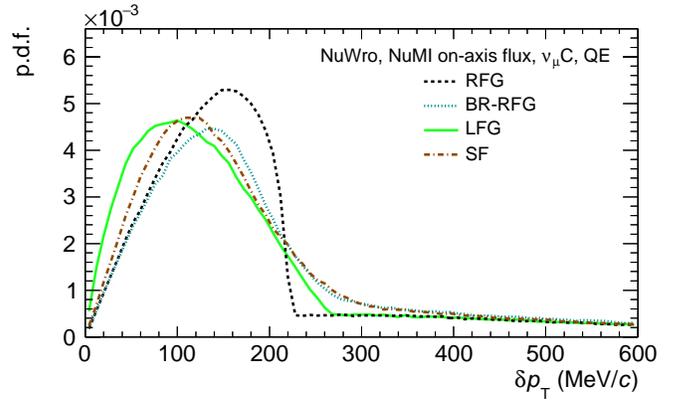}
\caption{NuWro predictions for $\dpt$ with different nuclear states: relativistic Fermi gas (RFG)~\cite{Moniz:1969sr}, relativistic Fermi gas with the Bodek-Ritchie modifications (BR-RFG)~\cite{Bodek:1980ar}, local Fermi gas (LFG)~\cite{Leitner:2008ue} and spectral function (SF)~\cite{Benhar:2006wy}. The NuMI~\cite{Anderson:1998zza} on-axis flux shape is used to simulate the neutrino energy distribution.}\label{fig:dpt_QE}
\end{center}
\end{figure}

The NuWro prediction for  $\dpt$ in QE  is shown in Fig.~\ref{fig:dpt_QE}. Four models of the nuclear state--- relativistic Fermi gas (RFG)~\cite{Moniz:1969sr}, relativistic Fermi gas with the Bodek-Ritchie modifications (BR-RFG)~\cite{Bodek:1980ar}, local Fermi gas (LFG)~\cite{Leitner:2008ue} and spectral function (SF)~\cite{Benhar:2006wy}---are compared. The deformation of the  $\ptnim$ shape due to FSI, which results in the long tail towards the upper end of the $\dpt$ distribution, is limited by the FSI strength quantified by $\tfsi$. For finite $\tfsi$, as is the case predicted by NuWro  (see e.g. Fig.~\ref{eq:ccintele}), the $\dpt$ shapes largely preserve the  Fermi motion distributions---a useful technique for understanding novel target materials in future experiments such as DUNE~\cite{Adams:2013qkq}. 

\begin{figure}
\begin{center}
\includegraphics[width=\columnwidth]{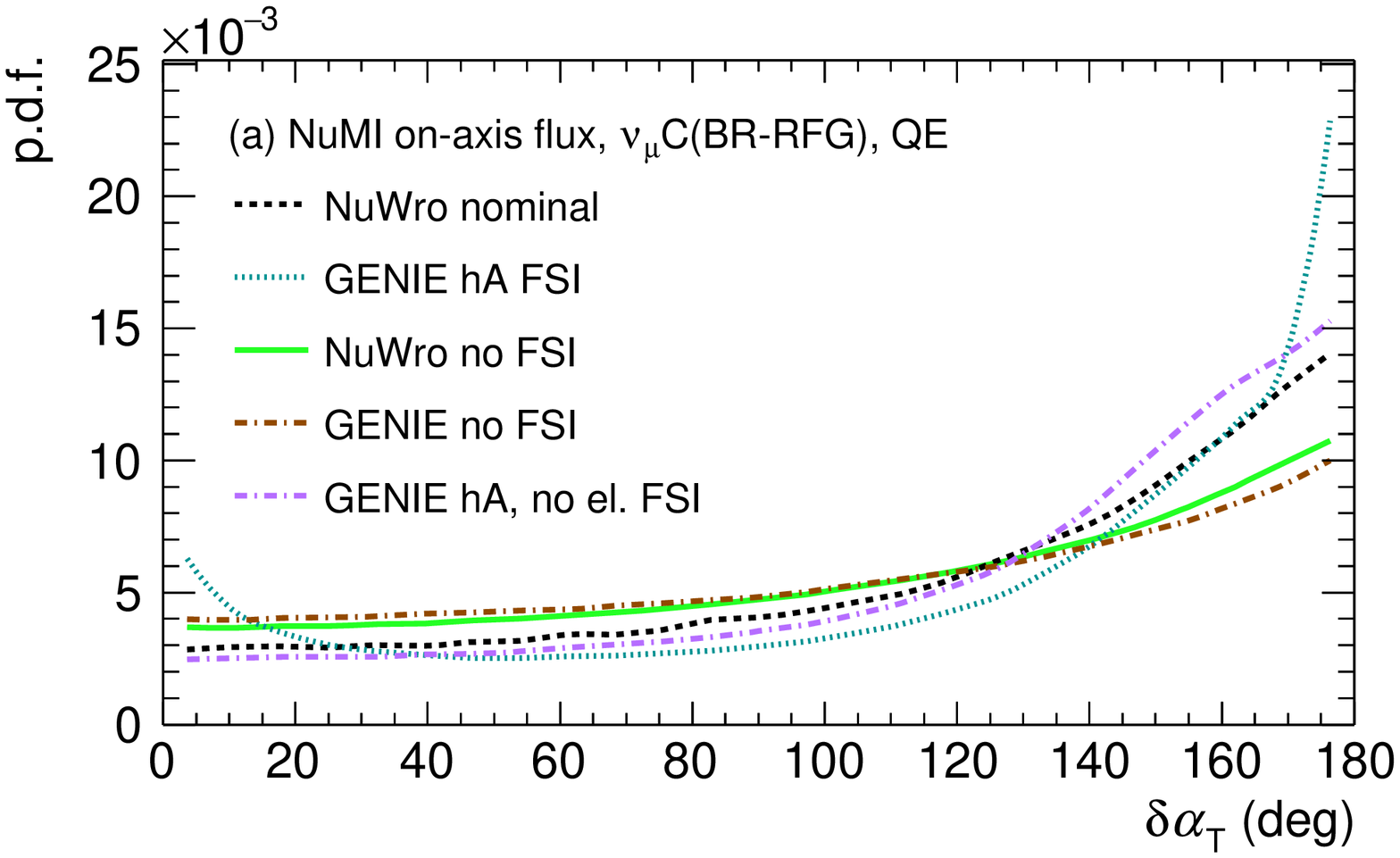}
\includegraphics[width=\columnwidth]{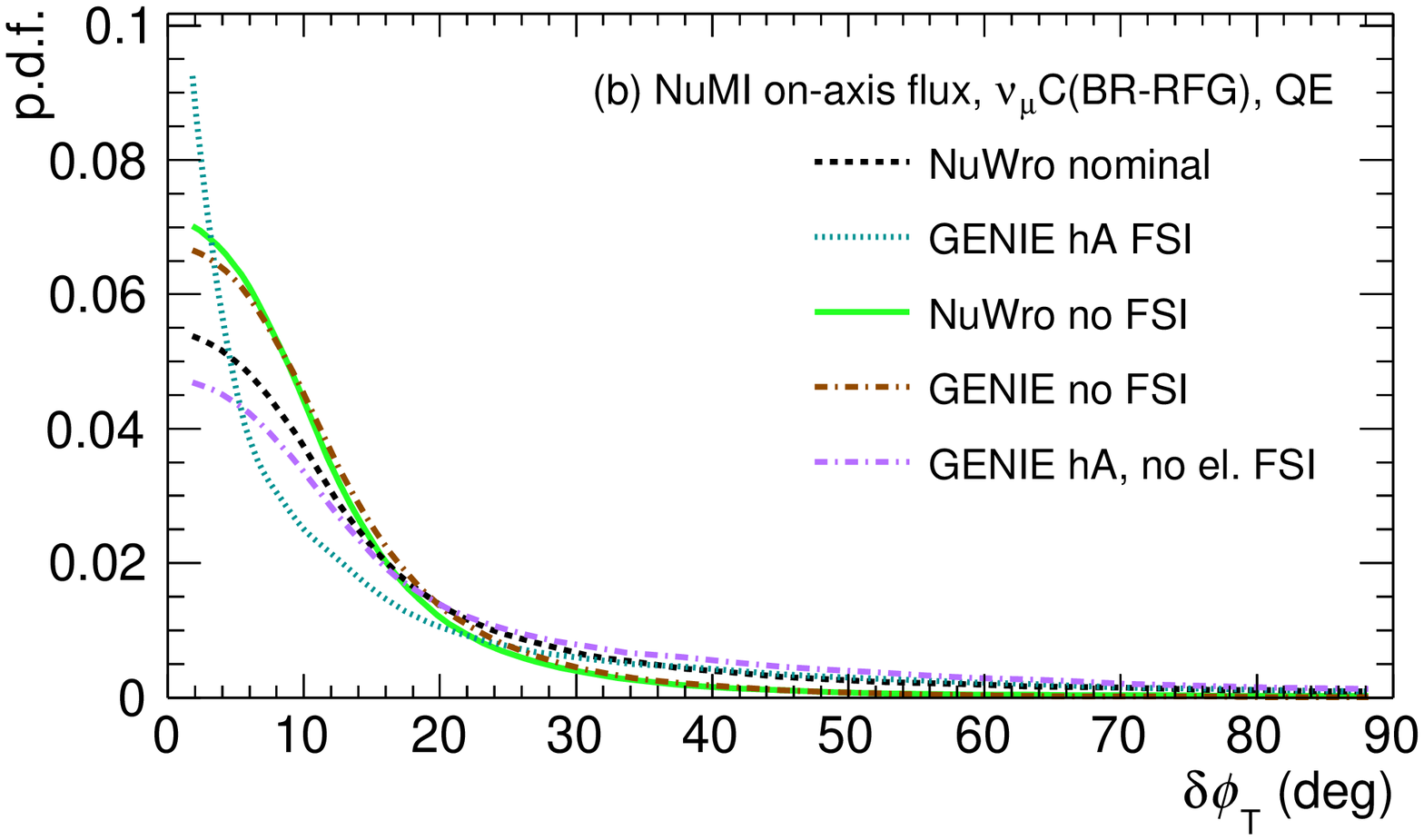}
\caption{NuWro and GENIE predictions for $\dalphat$ (upper) and
    $\dphit$ (lower). Nominal distributions are compared to the cases where FSI is disabled. Further comparison is made by removing nominal GENIE events that experienced proton elastic FSI (see text for exact definition). }\label{fig:neutrino_dat_dphit}
\end{center}
\end{figure}

The NuWro and GENIE predictions for $\dalphat$ and $\dphit$ in QE are shown in Fig.~\ref{fig:neutrino_dat_dphit}. When FSI is switched off in both simulations, consistent distributions are observed.  With the nominal settings, the two predictions significantly differ in the $\qt$-collinear regions---$\dalphat\sim$ 0, 180 degrees and $\dphit\sim0$ degrees---where GENIE predicts a much enhanced probability.  While the NuWro distributions show \emph{normal} evolution when the FSI is switched on as one would expect from the in-medium deflection and deceleration caused by FSI, the GENIE distributions show an \emph{inverted} tendency. Motivated by this observation, the GENIE Collaboration suggested to investigate the effect of the \emph{elastic interaction}  of  the hA FSI model. In the nominal GENIE simulation for QE on carbon,   events with protons that undergo this FSI interaction amount to about 40\% at the NuMI beam energy. After removing these events,  the GENIE prediction is more consistent with the NuWro nominal one, as is shown in Fig.~\ref{fig:neutrino_dat_dphit}. Further investigation taking into account the dependence on $\ptlm$ (in a similar approach to Fig.~\ref{fig:2ddat}) shows that in $\dalphat$ the collinear enhancement is of an appearant acceleration feature at low $\qtm$ ($\lesssim200$~MeV/$c$ at NuMI energy), and deceleration at high $\qtm$.

\section{Discussion} 

\begin{figure}
\begin{center}
\includegraphics[width=\columnwidth]{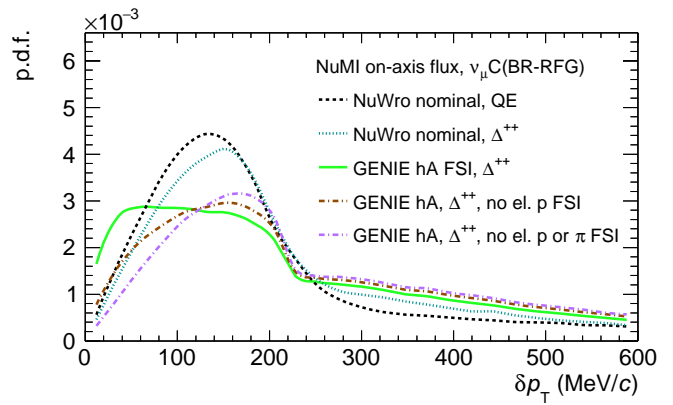}
\caption{NuWro and GENIE predictions for $\dpt$ in QE and $\Delta^{++}$ (p$\pi^+$) production. Nominal GENIE events that experienced proton and pion elastic FSI are removed stepwise to separate the effects; similar features are also exhibited in the p$\pi^-$ and p$\pi^0$ channels.}\label{fig:neutrinoRes_dpt}
\end{center}
\end{figure}

The definitions of the transverse kinematic imbalance require an exclusive measurement of the primary final-state particles. 
In RES, the imbalance is defined between the charged lepton and the $\proton\pi^{+(-)}$-system  with a $\nu$ ($\bar{\nu}$) beam. Transverse kinematic imbalance in  RES, proposed for the first time, should provide information on the resonance and pion FSIs. As an example, the NuWro and GENIE predictions for $\dpt$ in $\Delta^{++}$ (p$\pi^+$) production is shown in Fig.~\ref{fig:neutrinoRes_dpt}.  In the NuWro predictions, the deformation of the $\ptnim$ shape is severer in  RES than in QE because of the additional FSI from the pion final state; for GENIE, both the proton and pion hA elastic interaction  contribute to an \emph{inverted} deformation of the $\ptnim$ shape towards the lower end of the $\dpt$ distribution. Because all simulations use the same nuclear state (BR-RFG), a stronger FSI in GENIE can be inferred by its more pronunced upper tail in the distribution. 

Another interesting example from the transverse kinematic imbalance in  RES is the Pauli blocking of the resonance decay product. 
If it does not affect the resonance momentum---this is expected since  the polarization of the decay product alone can vary to fulfill Pauli blocking---the $\dalphat$ distribution at small $\ptlm$ ($\lesssim\pF$) will not be suppressed at 180 degrees, different from the QE case.

So far the discussion only considers the in-medium energy-momentum transfer and nuclear emission that are induced by FSIs. 
In reality, multinucleon correlations in the initial state could have significant impact in both processes 
that are however non-distinguishable from being FSI-induced on an event-by-event basis.    Their  contribution to the kinematic imbalance results in an additional convolution to each of the single transverse variables. The $\dpt$ and $\dalphat$ distributions---especially in the regions beyond the Fermi momentum and around the backward  peak, respectively---may therefore provide a kinematic constraint on the multinucleon correlations. Furthermore, because the nuclear emission probability is enhanced, its correlations   against the single transverse variables may be of  interest to provide sensitive signatures.

In general,   samples of QE and RES can be produced by selecting events with  at least one  proton and no pion, and at least  one proton and one pion, respectively, in a CC inclusive sample.  Neutrino energy-dependent contamination comes from other channels where incomplete final states are selected. In addition, because of their intrinsic kinematic imbalance, such non-exclusive backgrounds mimic the exclusive signals that experience FSIs and/or multinucleon correlations. However, quantitatively they may behavior differently: because of the characteristic strengths of the imbalance, one would expect that the single-transverse variables from the  background, the signal with FSIs,  and that with multinucleon correlations form a certain ordering, based on which it might be possible to perform efficient  channel purification and nuclear effect identification.

The observation of the transverse kinematic imbalance relies on the momentum reconstruction of both leptonic and hadronic final states. Since the $\dptv$ distributions reveal nuclear effects including Fermi motion,  deflection and acceleration/deceleration in FSI,  efforts in improving the momentum  and angular resolution, as well as  in understanding the momentum scales of the final-state lepton, proton and pion  might be needed for current experiemnts. Experiments with trackers, for example T2K~\cite{Abe:2011ks} and MINER$\nu$A~\cite{Aliaga:2013uqz}, might be capable to perform pioneering measurements to validate the different model predictions presented in the previous section. The prospect in  MINER$\nu$A  with its multiple-type nuclear targets  is of special interest because  the transverse kinematic imbalance would directly show the evolution of the nuclear effects as the target type varies. 

\begin{figure}
\begin{center}
\includegraphics[width=\columnwidth]{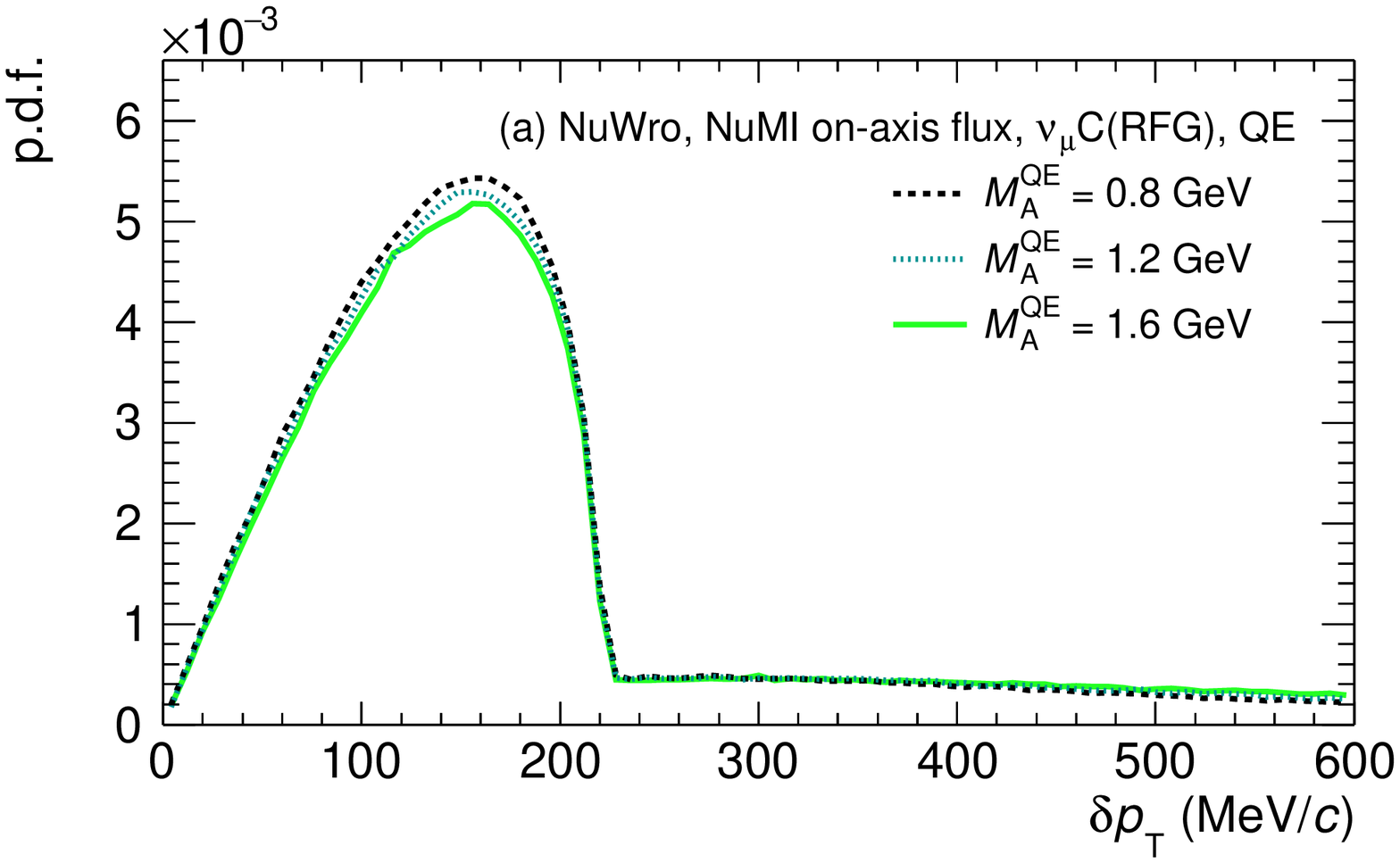}
\includegraphics[width=\columnwidth]{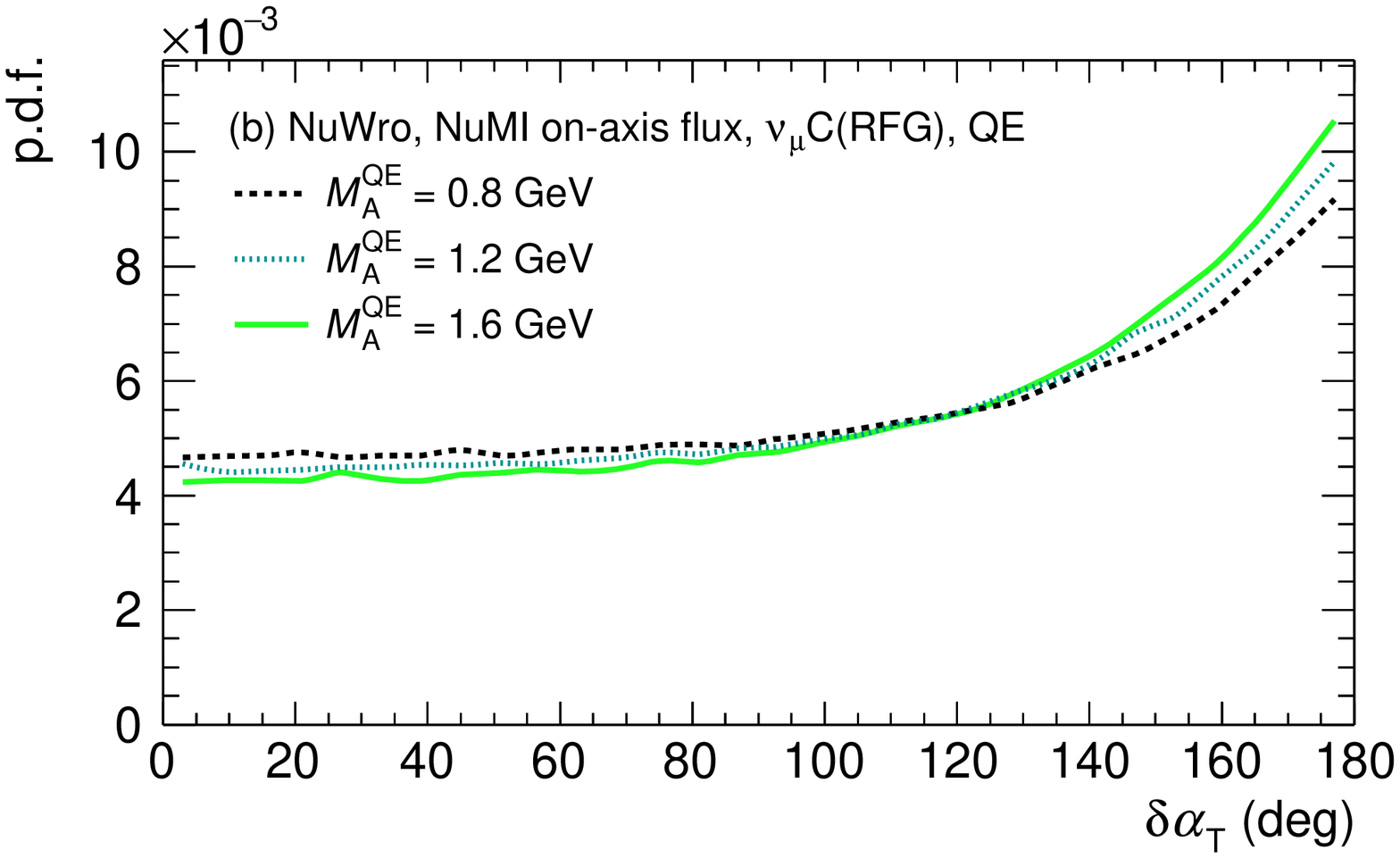}
\includegraphics[width=\columnwidth]{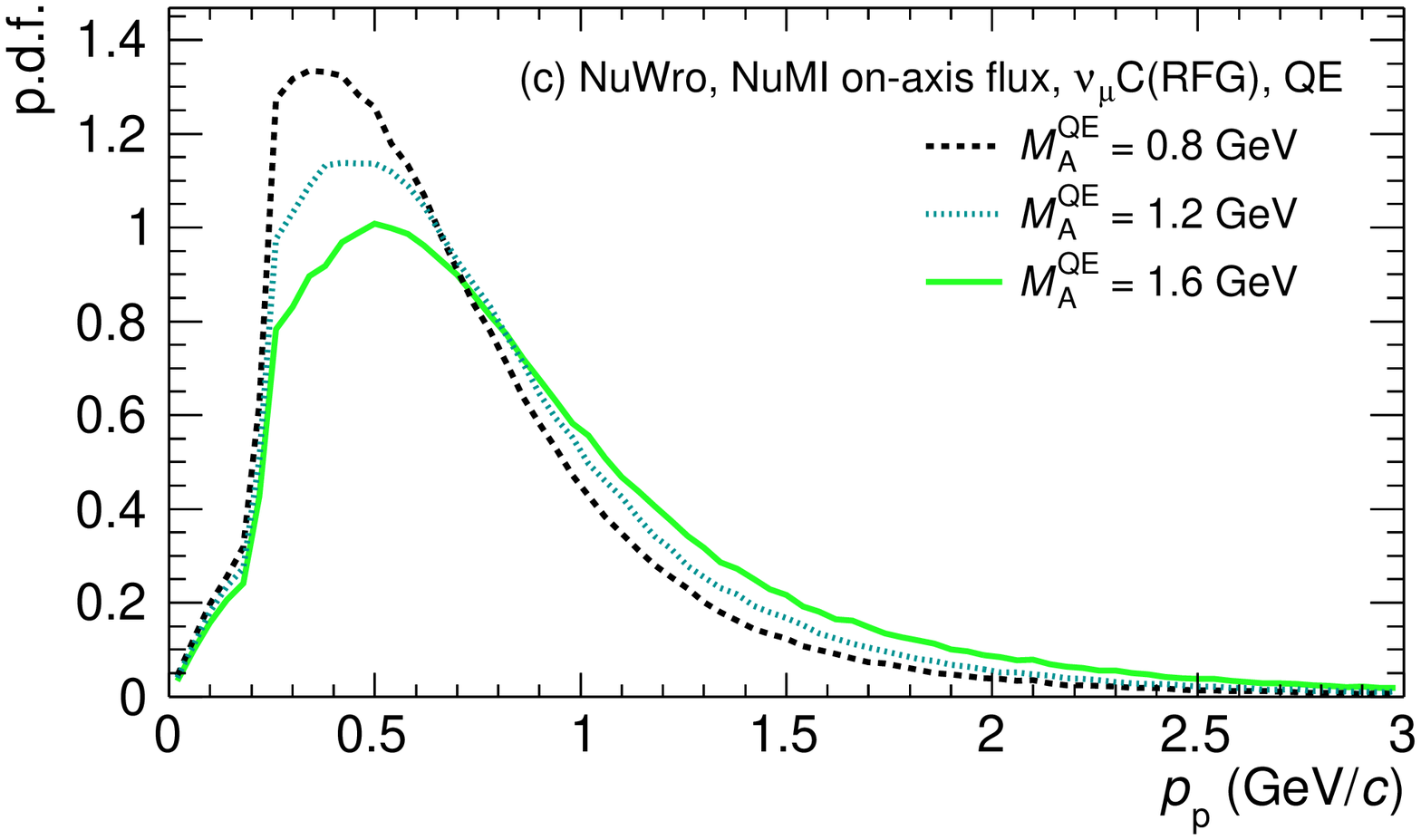}
\caption{Comparison of  $\dpt$ (upper), $\dalphat$ (middle) and the final-state proton momentum $p_\proton$ (lower) in QE for different values of the axial mass $\maqe$: 0.8, 1.2 (nominal) and 1.6 GeV. It shows that the transverse kinematic imbalances are much less sensitive to the variation of $\maqe$.}\label{fig:maqe}
\end{center}
\end{figure}

Throughout this work, it is emphasized that the equivalence between nuclear effects and transverse kinematic imbalance is invariant with the neutrino energy. In fact this equivalence is invariant with \emph{all} physics at the neutrino-nucleon interaction level. In analogue to the minimal neutrino energy dependence, transverse kinematic imbalances are also minimally influenced by nucleon level model uncertainties, such as the details of the nucleon form factors (Fig.~\ref{fig:maqe}). This further reduces the uncertainties in measuring nuclear effects.

Even though this work is motivated for neutrino interactions, the technique can be used in electron-nucleus scattering to probe  nuclear effects in general.
 In such a case, as the electron beam energy is well known, the energy imbalance between the initial and final states can also be studied. 

\section{Summary} 

In this work, for a  measurement of  nuclear effects independent of the neutrino energy, transverse kinematic imbalance  has been systematically examined. Its novel application  to the resonance production is important for  experiments with high energy neutrinos produced in the NuMI and LBNF~\cite{Adams:2013qkq} beam lines because of the significant production cross section;  it is also unique for the  study of  nuclear effects in antineutrino interactions where the final-state neutrons in quasielastic scattering cannot usually be  measured. 
 An extension of the single-transverse kinematics in the resonance production is the double-transverse momentum imbalance~\cite{Lu:2015hea, Lu:2015vri}, which is defined on the  axis perpendicular to both the neutrino and charged lepton momenta, with a sensitivity to the difference between the proton and pion FSIs. A  comprehensive description of nuclear effects in neutrino interactions should be attainable from  future measurements of the  single- and double-transverse kinematic imbalance on various nuclear targets in different interaction channels.

\section*{Acknowledgment}

The authors express their gratitude to the NuWro  and  GENIE Collaborations for the helpful discussions about using the generators in this work, and to K.~Duffy, S.~Dytman, Y.~Hayato,  K.~McFarland, R.~Shah, J.~Sobczyk, T.~Stewart, and C.~Wilkinson for helpful discussions. This work is supported by the UK Science and Technology Facilities Council, and (TY) by DOE and the DOE Early Career program, USA. 



\begin{thebibliography}{00}
\bibitem{binding}
The full form with the intra-nucleus potential will modify the apparent mass and the relativistic energy-momentum conversion of the initial nucleon.
\bibitem{polbias}
The bias to the mean $\omega$ is about 1\% in a NuWro  estimation (for technical detail see caption of Fig.~\ref{fig:pneqecmp}).
\bibitem{noabs}
The exclusive measurement discussed in this work is not sensitive to  the absorption and charge exchange in FSIs, and therefore these effects are not considered.
\bibitem{tausf}
The in-medium interaction probability is a functional of the nucleon momentum distribution in the nucleus.
\bibitem{deexcit} 
For simplicity, the nucleus excitation  during the basic interaction is not discussed here.
\bibitem{pnedef}
To be precise, the nuclear emission considered here is from the deexcitation within a detection time window. In practice,   slow emission with a longer transition time  is effectively seen as no emission. Another way the in-medium energy-momentum transfer is carried away is via the nucleus recoil.
\bibitem{gendep}
Conclusions in this work are independent of the chosen generator for demonstration.
\end{thebibliography}
\end{document}